\documentclass[%
 aip,
 jmp,%
 amsmath,amssymb,
 reprint,%
]{revtex4-1}

\usepackage{bibentry}
\usepackage{graphicx}
\usepackage{dcolumn}
\usepackage{bm}

\begin{document}

\preprint{AIP/123-QED}

\title{Magnetic field generation from composition gradients in inertial confinement fusion fuel}

\author{James D. Sadler}
\affiliation{%
Los Alamos National Laboratory, P.O. Box 1663, Los Alamos, NM 87545, USA}%
\author {Hui Li}
\affiliation{%
Los Alamos National Laboratory, P.O. Box 1663, Los Alamos, NM 87545, USA}%
\author {Kirk A. Flippo}
\affiliation{%
Los Alamos National Laboratory, P.O. Box 1663, Los Alamos, NM 87545, USA}%

\date{\today}

\maketitle

\textbf{
    Experimental asymmetries in fusion implosions can lead to magnetic field generation in the hot plasma core. For typical parameters, previous studies found that the magnetisation Hall parameter, given by the product of the electron gyro-frequency and Coulomb collision time, can exceed one. This will affect the hydrodynamics through inhibition and deflection of the electron heat flux. The magnetic field source is the collisionless Biermann term, which arises from the Debye shielding potential in electron pressure gradients. We show that there is an additional source term due to the $Z$ dependence of the Coulomb collision operator. If there are ion composition gradients, such as jets of carbon ablator mix entering the hot-spot, this source term can rapidly exceed the Biermann fields. In addition, the Biermann fields are enhanced due to the increased temperature gradients from carbon radiative cooling. With even stronger self-generated fields, heat loss to the carbon regions will be reduced, potentially reducing the negative effect of carbon mix.}

\section{Introduction}
 Inertial confinement fusion (ICF) experiments have reached the alpha heating regime, in which energy from fusion products is a significant contributor to the fuel energy balance and almost exceeds the radiative and conduction losses. The experimental Lawson parameter, given by the areal density and temperature product, is within $30\,$\% of the expected ignition threshold \cite{1}. If ignition is achieved, the fuel will rapidly self heat on a picosecond timescale and increase the total yield by more than a factor of 100 over current experiments. This is because the fusion rate is a strong function of temperature. A significant fraction of the milligrams of deuterium-tritium fuel will react, giving a megajoule scale yield.
 
 As the ignition threshold gets nearer, other physical processes will become more important. One important aspect is the self-generation of magnetic fields during the implosion. These fields occur due to the Biermann battery mechanism of magneto-hydrodynamics (MHD), and tend to wrap azimuthally around any intrusive plasma deformities. One study found that the rapid growth rates and radial compression can cause field strengths to approach $10^4\,$T, an exceedingly large value \cite{2}. This is high enough that it will indirectly affect hydrodynamics by inhibiting and deflecting the electron heat conduction. 
 
 In this work, we show that the $\mathbf{B}$ field may be even larger than previously thought, since there is an additional collisional magnetic source term. We derive this thermo-electric mechanism, discuss its physical origin and compare its magnitude to the Biermann term. The new term acts on ion composition gradients, such as those found at the edge of the carbon mix jets entering the hot-spot. Furthermore, the term scales with temperature, meaning that the field production will be extremely rapid in fusion conditions. We also discuss enhancement of the Biermann term in carbon mix regions due to the greater radiative cooling increasing the hot-spot temperature gradients. 
 
 Jets of carbon ablator mix have been measured entering the fuel hot-spot, with a typical total mass of up to $100\,$ng \cite{3, 4}. This mixing has a detrimental effect because the Bremsstrahlung radiative rate increases with the ion charge state. The radiation escapes, meaning the carbon region reaches a cooler temperature than the rest of the hot-spot. It then acts as a heat sink, with little fusion occurring within the mix region but a large amount of alpha particle and electron heat conduction into it. This energy is rapidly radiated away, with a measured loss of overall fusion yield \cite{3}. The magnetic insulation effect could reduce the detrimental heat loss into these mix regions.
 
 The magneto-hydrodynamics model is expected to be valid for national ignition facility deuterium-tritium hot-spot conditions, which have typical temperature $5\,$keV, density $100\,$gcm$^{-3}$, radius $30\,\mu$m and areal density $0.3\,$gcm$^{-2}$. Under these hot-spot conditions, the Coulomb logarithm $\ln(\Lambda)$ is in the range 2 to 5, sufficiently high that the light elements composing the hot-spot will be fully ionised and the classical transport coefficients should be valid. In addition, the Debye length $\lambda_D\simeq 10^{-10}\,$m is significantly shorter than any plasma scale-lengths, allowing the quasi-neutral approximation. In terms of the electron mass $m_e$, charge $e$, number density $n_e$ and temperature $T_e$, average ion charge state $\tilde Z=(\sum_jZ_j^2n_j)/(\sum_jZ_jn_j)$ and vacuum permittivity $\epsilon_0$, the corresponding electron-ion coulomb collision time 
\begin{align}
\tau &= \sqrt{\frac{9\pi}{2}}\frac{4\pi\epsilon_0^2\sqrt{m_e}T_e^{3/2}}{n_e\tilde Z e^4\ln(\Lambda)}\label{tau}
\end{align}
is approximately $1\,$fs. This leads to a mean free path of $\lambda_e=40\,$nm and Knudsen number $\lambda_e|\nabla T_e|/T_e \simeq 0.005$. Since the electron and ion mean free paths are much less than the gradient scale-lengths, the kinetic non-local corrections to the heat flux and fusion reactivity will be minimal \cite{5, 6, 7}. This also ensures that the MHD fluid approximation is valid. 

\section{Derivation}
The Braginskii generalised Ohm's law gives the steady-state electric field, including the effects of magnetised Coulomb collisions. The collisional behaviour depends on the dimensionless magnetisation parameter $\chi=\omega\tau=e|\mathbf{B}|\tau/m_e$. The plasma electric field is \cite{8}
\begin{align}
    \mathbf{E} &= -\mathbf{u}\times \mathbf{B} + \frac{\mathbf{J\times B}}{n_ee} - \frac{\nabla p_e}{n_ee} + \underline\eta.\mathbf{J} - \frac{1}{e}\underline\beta.\nabla T_e.\label{ohm}
\end{align}

We have neglected the terms due to electron inertia and inter-species ion diffusion, since in sub-sonic hot-spot conditions these are smaller by the electron-ion mass ratio. The ideal term $-\mathbf{u\times B}$ is due to the relativistic transform from the fluid frame, at fluid velocity $\mathbf{u}$, to the laboratory frame. The Hall term gives the effects of currents $\mathbf{J}$. The third term gives the Debye shielded potential, occurring because the electron pressure $p_e$ must be counteracted by a charge imbalance with an electric potential. 

The final two terms are due to the Coulomb collision operator. The resistive term is fairly intuitive, in that electrons carrying current will be scattered randomly by collisions with ions on a timescale $\tau$, neutralising the current. In terms of the dimensionless transport coefficients $\alpha_\perp(\tilde Z, \chi)$, $\alpha_\wedge(\tilde Z, \chi)$ and $\alpha_0(\tilde Z) = \alpha_\perp(\tilde Z, 0)$, the full tensor form is given by \cite{8}
\begin{align}
    \underline\eta.\mathbf{J} = \frac{m_e}{n_ee^2\tau}\left(\alpha_0\mathbf{\hat b}(\mathbf{J.\hat b}) + \alpha_\perp\mathbf{\hat b\times(J\times\hat b)} - \alpha_\wedge\mathbf{\hat b\times J }\right)\label{eta}
\end{align}

The orthogonal basis vectors are given in terms of the magnetic field direction $\mathbf{\hat b = B/|B|}$. The $\alpha_0$ term is independent of $\chi$, since transport along the field direction cannot be affected by magnetic fields. The second term gives the resistive electric field across the field lines. The perpendicular resistivity coefficient $\alpha_\perp$ increases as $\chi$ increases. The magnetic deflection also introduces a third term which is perpendicular to both the field and the driving current. This off-diagonal term is maximal for around $\chi\simeq 5$. The transport coefficients must be found numerically from the Vlasov-Fokker-Planck equation. Fits to the dimensionless coefficients are given in reference \cite{8}.

Similarly, the collisional thermal force is given in terms of the dimensionless transport coefficients $\beta_\perp(\tilde Z, \chi)$, $\beta_\wedge(\tilde Z, \chi)$ and $\beta_0(\tilde Z) = \beta_\perp(\tilde Z, 0)$ by
\begin{align}
    \underline\beta.\nabla T_e = \beta_0\mathbf{\hat b}(\nabla T_e\mathbf{.\hat b}) + \beta_\perp\mathbf{\hat b\times}(\nabla T_e\mathbf{\times\hat b)} + \beta_\wedge\mathbf{\hat b\times  }\nabla T_e\label{beta}.
\end{align}
The collisional thermal force is due to the electron velocity dependence of the coulomb Collision rate. It arises because, even in pressure equilibrium, if there is a temperature gradient then faster electrons from the hotter side will be less collisional with the ions [eq. (\ref{tau})]. This means there is a net force on the electrons towards the colder side, which is balanced by an electric field also towards the colder side. 

\begin{figure*}[t]
  \includegraphics{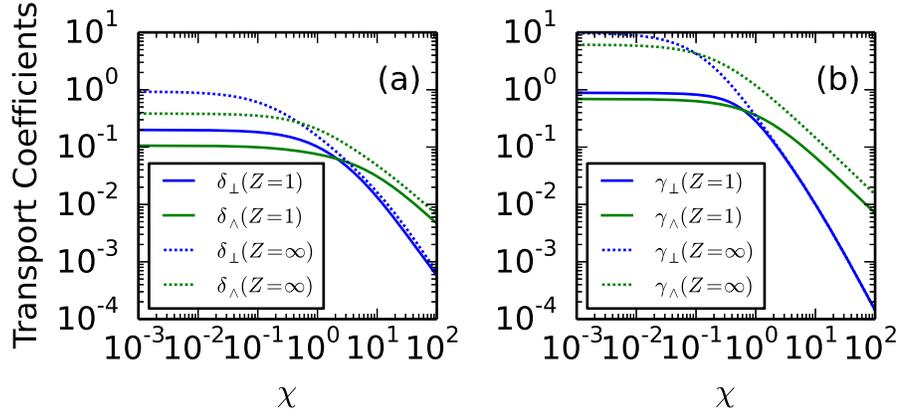}
  \caption{Plots of the extended-MHD $\delta$ and $\gamma$ transport coefficients, giving the effect of the extended-MHD collisional terms on the magnetic field advection velocity. Both are shown for ion charge state $Z=1$ and in the limit for $Z\rightarrow\infty$. (a) The Hall velocity correction coefficients. (b) The Nernst velocity and cross-gradient Nernst velocity coefficients.}
   \label{coefs}
\end{figure*}

The $\mathbf{J}$ and $\nabla T_e$ vectors can be decomposed into their components parallel and perpendicular to the field, via the identity $\mathbf{J}=\mathbf{\hat b}(\mathbf{J.\hat b}) + \mathbf{\hat b\times}(\mathbf{J\times\hat b)}$. The resistive term can then be manipulated to give
\begin{align}
    &\underline\eta.\mathbf{J} = \frac{m_e}{n_ee^2\tau}\left(\alpha_0\mathbf{J} + (\alpha_\perp-\alpha_0)\mathbf{\hat b\times(J\times\hat b)} - \alpha_\wedge\mathbf{\hat b\times J }\right)\\
    \begin{split}
  &=  \frac{m_e}{n_ee^2\tau}\left[ \alpha_0\mathbf{J} + \frac{\mathbf{B}}{\mathbf{|B|}}\times\left( -\alpha_\wedge\mathbf{J } + (\alpha_\perp-\alpha_0)\mathbf{(J\times\hat b)}\right)\right].\label{eta2} \end{split}
\end{align}
We make the standard MHD approximation to neglect the displacement current in the Maxwell equations, effectively eliminating high frequency oscillation modes and electron waves, giving $\mathbf{J} = c^2\epsilon_0\nabla\times\mathbf{B}$. We also use the definition $\chi=e|\mathbf{B}|\tau/m_e$ of the magnetization and define the magnetic diffusivity $\eta_0=m_ec^2\epsilon_0\alpha_0/(n_ee^2\tau)$, to give

\begin{align}
        \underline\eta.\mathbf{J} &=  \eta_0\nabla\times\mathbf{B} - \mathbf{u_\alpha\times B},\label{alpha}\\
        \mathbf{u_\alpha} &= \frac{1}{n_ee}\left[-\delta_\perp\mathbf{J} + \delta_\wedge(\mathbf{J\times\hat b})\right],
\end{align}
where, following reference \cite{9}, we have also defined the Hall velocity correction coefficients $\delta_\perp(\tilde Z, \chi) = \alpha_\wedge/\chi$ and $\delta_\wedge(\tilde Z, \chi) = (\alpha_\perp-\alpha_0)/\chi$. These coefficients are plotted in Fig. \ref{coefs}a. They are dimensionless and positive for all $\chi$ and $\tilde Z$. Comparing eq. (\ref{alpha}) to eq. (\ref{ohm}), it is clear that the collisional resistance alters the advection velocity of the magnetic field, with a term of the same functional form as $-\mathbf{u\times B}$.

The thermoelectric term $-\frac{1}{e}\underline\beta.\nabla T_e$ can be similarly decomposed to give

\begin{align}
      &-\frac{1}{e}\left(\beta_0\nabla T_e + (\beta_\perp-\beta_0)\mathbf{\hat b\times(\nabla} T_e\mathbf{\times\hat b)} + \beta_\wedge\mathbf{\hat b\times }\nabla T_e\right)\\
  =  &-\frac{\beta_0}{e}\nabla T_e + \frac{\mathbf{B}}{e|\mathbf{B}|}\times\left[ -\beta_\wedge\nabla T_e + (\beta_0-\beta_\perp)(\nabla T_e\times\mathbf{\hat b)}\right]\label{beta2}.
\end{align}
Again using $\chi=e|\mathbf{B}|\tau/m_e$ and defining the Nernst velocity coefficient $\gamma_\perp(\tilde Z, \chi) = \beta_\wedge/\chi$ and the cross-gradient Nernst coefficient $\gamma_\wedge(\tilde Z, \chi) = (\beta_0-\beta_\perp)/\chi$, the thermoelectric contribution to the electric field can be written \cite{9}
\begin{align}
        -\frac{1}{e}\underline\beta.\nabla T_e &= -\frac{1}{e}\beta_0\nabla T_e -  \mathbf{u_\beta\times B}\\
        \mathbf{u_\beta} &= \frac{\tau}{m_e}\left[ -\gamma_\perp\nabla T_e + \gamma_\wedge(\nabla T_e\times\mathbf{\hat b})\right].
\end{align}
Similarly to the $\delta$ coefficients, the newly defined Nernst coefficients are dimensionless, positive and tend towards finite order 1 values for low magnetization. The $\delta(\tilde Z, \chi)$ and $\gamma(\tilde Z, \chi)$ coefficients are plotted in Fig. \ref{coefs} for $\tilde Z=1$ and $\tilde Z\rightarrow\infty$. Note that these coefficients have been calculated using the fit functions in reference \cite{8}, which can lead to inaccuracies in the cross-gradient coefficients in the limit of low magnetization. Physically, they should tend to zero for low magnetization. More accurate fits will be explored in future work.

The total extended-MHD electric field can therefore be written in the form
\begin{align}
    \mathbf{E} &= -\mathbf{u_B\times B} -\frac{\nabla p_e}{n_ee} + \eta_0\nabla\times\mathbf{B} - \frac{1}{e}\beta_0\nabla T_e,\label{totalohm}
\end{align}
where the total field advection velocity $\mathbf{u_B}$ has been altered by the Coulomb collisions and will be discussed in the following section.

\section{Discussion}
To see the magnetic field evolution, eq. (\ref{totalohm}) can be substituted into the Maxwell equation $\partial_t\mathbf{B} = -\mathbf{\nabla\times E}$. Using the ideal gas equation of state $p_e=n_eT_e$, the pressure gradient term yields the Biermann battery magnetic source term

\begin{align}
  \frac{\partial\mathbf{B}}{\partial t} = \nabla\times\left(\frac{\nabla p_e}{n_ee}\right) = -\frac{\nabla n_e\times \nabla p_e}{n_e^2e} = -\frac{\nabla n_e\times \nabla T_e}{n_ee}.
\end{align}

The resistive term can be simplified using the identity $-\nabla\times(\eta_0\nabla\times\mathbf{B}) = \eta_0\nabla^2\mathbf{B} - \nabla\eta_0\times(\nabla\times\mathbf{B})$ with $\mathbf{\nabla.B}=0$. The thermoelectric term can be simplified with the same identity, along with the fact that $\nabla\times\nabla T_e = 0$.

The final form of the induction equation is therefore composed only of an advection term, a diffusion term, the resistivity gradient term and two source terms that are still active even when $\mathbf{B}=0$  \cite{9},

\begin{align}
\begin{split}
  \frac{\partial\mathbf{B}}{\partial t} =&\nabla\times(\mathbf{u_B}\times \mathbf{B}) + \eta_0\nabla^2\mathbf{B} - \nabla\eta_0\times(\nabla\times\mathbf{B})\\
  &-\frac{\nabla n_e\times\nabla T_e}{n_ee} + \frac{\beta_0^\prime(\tilde Z)}{e}\nabla\tilde Z\times\nabla T_e.\label{induction}
\end{split}
\end{align}
The first term causes advection of the magnetic field at velocity $\mathbf{u_B}$, although it has no effect when the advection is along the field line. The field advection velocity is given by
\begin{align}\begin{split}
    \mathbf{u_B} &= \,\mathbf{u} -(1+\delta_\perp)\frac{\mathbf{J}}{n_ee} + \delta_\wedge\frac{\mathbf{J\times\hat b}}{n_ee} \\&+\frac{\tau}{m_e}\left(-\gamma_\perp\nabla T_e + \gamma_\wedge\nabla T_e\times\mathbf{\hat b}\right)\label{ub}
\end{split}
\end{align}

It is now clear that the sole effect of the anisotropic $\perp$ and $\wedge$ extended-MHD  terms is to alter the magnetic field advection velocity. Instead of having $\mathbf{u_B} =\mathbf{u}$ as in ideal MHD, the advection velocity now also includes the the Hall velocity, with some small correction terms containing the $\delta$ coefficients. From Fig. \ref{coefs}, it is clear that, for $Z=1$, the Hall velocity corrections do not exceed $20\,\%$. The advection also includes the Nernst velocity from the thermoelectric term, which advects the field down electron temperature gradients at a speed similar to the flow of heat from electron conduction. Due to the large heat fluxes in fusion hot-spots, the Nernst advection can significantly alter the magnetic field profile. However, the Hall velocity terms (those containing $\mathbf{J}$) are typically small in ICF hot-spot conditions, on the order of $100\,$ms$^{-1}$. This is compared to $10^5\,$ms$^{-1}$ for the fluid and Nernst velocities. There is also the cross-gradient $\gamma_\wedge$ Nernst advection term, which advects the field along isotherms, in the direction of $\nabla T_e\times\mathbf{B}$.

Use of a non-zero resistivity causes a diffusion of the magnetic field, whose strength is characterised by the dimensionless magnetic Reynolds number $R_M = UL/\eta_0$, where $U$ is a typical velocity and $L$ is a typical length scale. For the hot-spot conditions, use of equation (\ref{tau}) gives $\eta_0 \simeq 10^{-2}$m$^2$s$^{-1}$. Taking $U\simeq 3\times 10^{5}$ms$^{-1}$ as a typical implosion velocity and $L$ as the hot-spot size, this gives $R_M\simeq 10^3$, meaning advection of the field is dominant over its diffusion and the $\eta_0$ terms are fairly small in the present case. The smoothing effect of the diffusion term over the stagnation time $t=100\,$ps can be estimated as $L_\mathrm{diff}=\sqrt{\eta_0t}\simeq 1\,\mu$m, giving a minimum length scale for the size of magnetic features.

It should be noted \cite{10} that when $\mathbf{J}$ is perpendicular to $\mathbf{B}$, the $\delta_\wedge$ advection term in eq. $(\ref{ub})$ is equivalent to additional diffusion of the magnetic field, such that the resistive terms in eq. (\ref{induction}) become $\eta_\perp\nabla^2\mathbf{B} - \nabla\eta_\perp\times(\nabla\times\mathbf{B})$, rather than $\eta_0\nabla^2\mathbf{B} - \nabla\eta_0\times(\nabla\times\mathbf{B})$. This is true, for example, in a two-dimensional geometry with self-generated fields. However, eq. (\ref{induction}) shows that the general formulation is that of isotropic diffusion with coefficient $\eta_0$, with the additional $\delta_\wedge$ advection term that will cause some additional anisotropic diffusion when $\mathbf{J}$ is not parallel to $\mathbf{B}$.

\begin{figure*}[t]
  \includegraphics{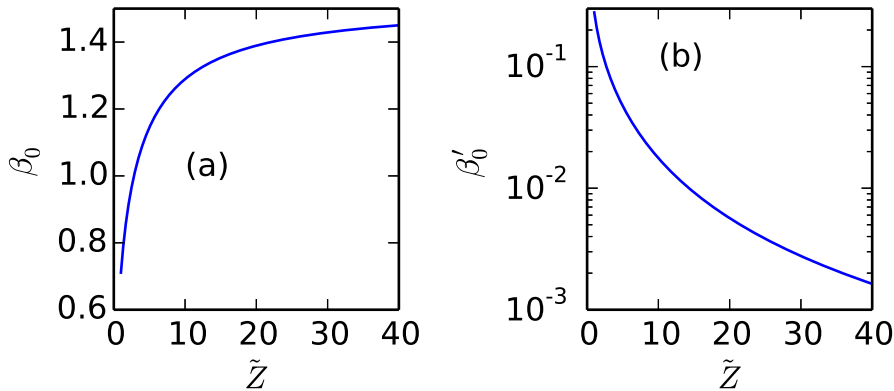}
  \caption{(a) Plot of the Braginskii thermo-electric coefficient $\beta_0(\tilde Z)$. (b) Plot of the derivative of $\beta_0$. Although the thermo-electric force increases for higher $Z$ plasmas, it is the derivative that gives the coefficient of the magnetic field production rate. This is maximal for low Z plasma such as DT fusion fuel, with $\beta_0^\prime(1)\simeq0.3$.}
   \label{betaplot}
\end{figure*}

The magnetic dynamics in inertial confinement fusion hot-spots are dominated by the advection term and the two source terms (final terms in eq. \ref{induction}). The Biermann term acts on misaligned density and temperature gradients, while the thermoelectric term acts on misaligned ion composition and temperature gradients. The quantity $\beta_0(Z)$ and its derivative $\beta_0'(Z)$ are plotted in Fig. \ref{betaplot}. Clearly the collisional source term will be maximal for low $\tilde Z$ plasmas with steep gradients in $\tilde Z$, whereas the Biermann term is independent of $\tilde Z$. 

In the context of inertial confinement fusion fuel impurities, there may exist a $\nabla \tilde Z$ due to carbon jets penetrating the burning fuel. The carbon region will reach equilibrium at a lower temperature than the rest of the hot-spot, since the Bremsstrahlung radiative losses increase with $\tilde Z$. This naturally introduces a $\nabla T_e$ away from the mix jet and a $\nabla \tilde Z$ towards the mix jet. Due to thermal conduction and hydrodynamic motion, these are unlikely to be exactly aligned. The conditions are therefore met for the collisional thermoelectric source term. 

The mix region will radiatively contract, leading to a $\nabla n_e$ towards the mix region and $\nabla T_e$ away from it. The magnitude of the Biermann term is then approximately $fT_e/(el_nl_T)$, where $l_T=T_e/|\nabla T_e|$ is the temperature gradient scale-length, $l_n=n_e/|\nabla n_e|$ is the density scale-length and $f=\sin\theta$ is a reduction factor due to the misalignment of the gradients. With typical hot-spot temperature $T_e=5\,$keV, $f=0.1$ and scale-lengths $3\,\mu$m, this gives field growth rate $50\,$Tps$^{-1}$. The field is thus expected to reach several thousand Tesla over the $100\,$ps
 stagnation time-scale. 
 
If carbon enters the hot-spot, the ion charge state gradient will be reduced by inter-species ion diffusion. This can be estimated using the model of Molvig, Simakov and Vold \cite{11}, in which an initially sharp interface between a light and heavy ion species will develop through diffusion. The diffusion coefficient can be estimated as $D=\frac{2T_i\tau}{Z^2\sqrt{m_im_e}}\simeq 0.03\,$m$^2$s$^{-1}$, similar to the thermal diffusion and resistive magnetic diffusion rates. Over the stagnation time $t=100\,$ps, this leads to a diffusive scale-length of $\sqrt{tD}=1.6\,\mu$m. This gives a lower bound on the expected scale-lengths $l_Z$, $l_n$ and $l_T$.

The collisional source term is maximal when $Z=1$, giving $\beta_0'(1)\simeq0.3$, meaning it is similar in magnitude to the Biermann term. For $Z\simeq 1$, as in the hydrogen hot-spot, this leads to field growth rate $0.3fT_e/(el_Zl_T)\simeq 20\,$Tps$^{-1}$. The collisional source term can be similar magnitude to the Biermann term. In fact, it may exceed it. This is because hydrodynamic motion acts to smooth the pressure gradients and reduce the Biermann growth. The collisional term has no such natural stabilisation, since it acts on composition gradients which can even exist in pressure equilibrium, such as in an ideal isobaric hot-spot. 

Carbon jets will also have increased Biermann fields relative to hydrogen jets. This is due to the increased radiative cooling providing a steeper temperature gradient in the carbon mix case. We also note that the two source terms are likely to be in opposite directions. This is because $\nabla n_e$ and $\nabla \tilde Z$ are both towards the centre of the mix region. However, $\beta_0^\prime$ is positive and the two terms have opposite signs [eq. (\ref{induction})], so will be in opposite directions. This may mean the magnetised hydrodynamics of hydrogen jets are quite different to that of carbon mix jets, since the alterations to heat flow could be in the opposite direction. For hydrogen jets, the magnetised Righi-Leduc heat-flow is towards the base of the jet \cite{2}. For carbon jets where the collisional thermoelectric source term is dominant, the magnetic field could be in the opposite direction and deflect heat towards the spike tip.

The Nernst advection must also be considered, since it will advect the magnetic field into the cooler mix region. Since the magnetisation scales as $T_e^{3/2}$, this will reduce the anisotropic heat flux effects. 

Another important consideration is the fusion product alpha particle transport. These alpha particles start with energy $3.5\,$MeV, giving a gyro-radius of $r_L=(|\mathbf{B}|/\mathrm{1000T})^{-1}\times270\,\mu$m. The estimates of the $5000\,$T field strength suggest that the minimal alpha gyro-motion could be close to the hot-spot size $r\simeq 30\mu$m. However, the scale-length of the magnetic field regions will be much smaller than this, meaning the alpha particle energy deposition profile will have only minor changes. The field strength would need to reach approximately $10^5\,$T for any appreciable magnetic confinement of the alpha particle energy.

Since most energy within the carbon regions is rapidly radiated away, heat flux into carbon regions is a primary loss mechanism from the plasma. Electron heat flux into the carbon region will be reduced by the magnetic field. This may help to insulate the carbon mix regions and slightly reduce their negative effects. 

In summary, the nature of the induction equation indicates that carbon impurities mixing into the fusion hot-spot may lead to larger magnetic fields than with hydrogen jets. The field generation rate is on the order of $50\,$Tps$^{-1}$. This increase is due to two mechanisms. Firstly, the temperature gradients around the spike will be larger due to the increased radiative cooling, leading to increased Biermann growth. Secondly, there is an additional collisional thermoelectric source of magnetic field that only occurs with gradients in the average ion charge state $\tilde Z$. This will only arise if higher Z impurities enter the hot-spot. These mechanisms will lead to magnetisation of the electron heat flux, affecting the hydrodynamics of the jet.

Research presented in this article was supported by the Laboratory Directed Research and Development program of Los Alamos National Laboratory under project number 20180040DR.




\end{document}